\documentstyle[aps,multicol,epsf]{revtex}
\begin{document}
\title{Interpretation of SAMPLE and HAPPEX Experimental Results on Strange
       Nucleon Form Factors}
\author{Stanislav Dubni\v cka\\
        Inst. of Physics, Slovak Academy of Sci., Bratislava, Slovak Republic\\
\medskip
        Anna Zuzana Dubni\v ckov\'a and Peter Weisenpacher\\
        Dept. of Theor. Physics, Comenius Univ., Bratislava, Slovak Republic}
\maketitle
\begin{abstract}
   A behaviour of strange nucleon form factors is predicted by means
of Jaffe's idea about the relations of $\omega$ and $\phi$ vector-meson
coupling constant ratios. Its application to a specific eight-resonance
unitary and analytic model of nucleon electromagnetic structure, describing
also the time-like nucleon electromagnetic form factor data, explains the
positive central values from recent SAMPLE and HAPPEX experiments.

\end{abstract}

\bigskip


\bigskip
\begin{multicols}{2}
   There has been considerable interest concerning the question of strangeness
contribution to the nucleon structure which is represented by various
strange operators, like $\bar{s} s$ extracted from the analysis of
the pion-nucleon $\Sigma$-term, $\bar{s} \gamma_{\mu} \gamma_{5} s$  measured
in deep-inelastic lepton scattering off protons and the vector current
$\bar{s} \gamma_{\mu} s$, accessible in parity-violating elastic and
quasielastic electron scattering from the proton and nuclei \cite{mus}.
More recently, a well-defined experimental program has just started
determination of the nucleon matrix element
$\langle{p'}|\bar{s}\gamma_{\mu} s|p\rangle$ of
the strange-quark vector current by means of the elastic scattering of
polarized electron beam on a liquid hydrogen target, in which unexpected
positive values of the strange nucleon form factors (FF's) were revealed.
The first result for the strange nucleon magnetic FF $G^{s}_{M}(t)$ at
$t=-0.1 GeV^2$ ($t=q^2=-Q^2$ is the four-momentum transfer squared)
associated with this matrix element has been reported by the SAMPLE
Collaboration at MIT/Bates Linear Accelerator Center \cite{mue}

\begin{equation}
    G^{s}_{M} (-0.1)=+0.23 \pm 0.37 \pm 0.15 \pm 0.19 [\mu_N].
\label{GM1}
\end{equation}
Later, a combination of the strange nucleon electric and
magnetic FF's at $t=-0.48 GeV^2$ has been determined by the HAPPEX
Collaboration at TJNAF \cite{ani}
\end{multicols}

\begin{equation}
 G^{s}_{E}(-0.48)+0.39 G^{s}_{M}(-0.48)=+0.023 \pm 0.034
\pm 0.022 \pm 0.026 .
\label{GEM}
\end{equation}

\begin{multicols}{2}
   As the error bars in both previous experimental results are large,
no strict conclusions about the strangeness in the nucleon could follow
from them.

   However, the SAMPLE Collaboration  at MIT/Bates Linear Accelerator Center
has just recently reported \cite{spa} a new experimental measurement of the
strange nucleon magnetic FF, in which an improved monitoring and control of
systematic errors as well as better statistical precision have been achieved.
As a result the clearly positive value of the strange nucleon magnetic FF at
$t=-0.1 GeV^2$ is found
\begin{equation}
   G^{s}_{M}(-0.1)=+0.61 \pm 0.17 \pm 0.21 \pm 0.19 [\mu_N]
\label{GM2}
\end{equation}
suggesting also the strange magnetic moment of the nucleon
$\mu_{s}$= $G^{s}_{M} (0)$ quite likely to be positive.

   In this Letter we present an explanation of these positive central
values by an application of Jaffe's idea \cite{jaf} to a specific
eight-resonance (four isoscalar and four isovector vector mesons) unitary
and analytic model of nucleon electromagnetic (EM) structure, by means of
which a behaviour of corresponding strange nucleon FF's is predicted.

   The momentum dependence of the nucleon matrix element of the strange-quark
vector current $J^{s}_{\mu}$=$\bar{s}\gamma_{\mu} s$ is contained in the Dirac
$F^{s}_{1} (t)$ and Pauli $F^{s}_{2} (t)$ strange nucleon FF's defined
by the relation

\begin{equation}
   \langle p^{'}|\bar{s}\gamma_{\mu} s|p\rangle=\bar{u}(p^{'})
\left[\gamma_{\mu} F^{s}_{1}+
i{{\sigma_{\mu\nu}q^\nu}\over{2m_N}} F^{s}_{2}\right]u(p)
\label{FFS}
\end{equation}
in a complete analogy to the nucleon matrix element

\begin{equation}
   \langle p^{'}|J^{I=0}_{\mu}|p\rangle=\bar{u}(p^{'})
\left[\gamma_{\mu} F^{I=0}_{1}+
i{{\sigma_{\mu\nu}q^{\nu}\over{2m_N}}} F^{I=0}_{2}\right]u(p)
\label{FF0}
\end{equation}
of the isoscalar EM current $J^{I=0}_{\mu}$={1/6}
$(\bar{u}\gamma_{\mu}u+\bar{d}\gamma_{\mu}d)$-{1/3}$\bar{s}
\gamma_{\mu}s$,
where $q^{\nu}=(p^{'}-p)^{\nu}$ is the four-momentum transfer,
$\bar{u}(p^{'})$, $u(p)$ are the free nucleon Dirac bi-spinors and
$F^{I=0}_{1}(t)$ and $F^{I=0}_{2}(t)$ are isoscalar parts of the Dirac and
Pauli nucleon EM FF's, respectively.

   We note here that (as a result of the isospin zero value of the strange
quark) the nucleon strange FF's $F^{s}_{1}(t)$ and $F^{s}_{2}(t)$ contribute
just to $F^{I=0}_{1}(t)$ and $F^{I=0}_{2}(t)$ and never to $F^{I=1}_{1}(t)$
and $F^{I=1}_{2}(t)$ which are defined by the matrix element

\begin{equation}
   \langle p^{'}|J^{I=1}_{\mu}|p\rangle=\bar{u}(p')
\left[\gamma_{\mu} F^{I=1}_{1}+
i{{\sigma_{\mu\nu}q^{\nu}\over{2m_N}}} F^{I=1}_{2}\right]u(p)
\label{FF1}
\end{equation}
of the isovector EM current $J^{I=1}_{\mu}$={1/2}$(\bar{u}
\gamma_{\mu}u-\bar{d}\gamma_{\mu}d)$.

   Then the main idea of a prediction of strange nucleon FF behaviours,
based on the $\omega-\phi$ mixing and on the assumption that the quark
current of some flavour couples with universal strength exclusively to the
vector-meson wave function component of the same flavour, consists in the
following.

   If one knows free parameters $(f^{(i)}_{\omega NN}/f^{e}_{\omega})$,
$(f^{(i)}_{\phi NN}/f^{e}_{\phi})$ (i=1,2) of the suitable model of
$F^{I=0}_{1}(t)$, $F^{I=0}_{2}(t)$ i.e.

\begin{equation}
  F^{I=0}_{i}(t)=f\left[t;({f^{(i)}_{\omega NN}}/{f^{e}_{\omega}}),
({f^{(i)}_{\phi NN}}/{f^{e}_{\phi}})\right] (i=1,2)
\label{Me}
\end{equation}
where $f^{(i)}_{\omega NN}, f^{(i)}_{\phi NN}$ are coupling constants of
$\omega$ and $\phi$ to nucleons and $f^{e}_{\omega}, f^{e}_{\phi}$ are
virtual photon$\leftrightarrow$V=$\omega,\phi$ coupling constants given
by the corresponding leptonic decay widths $\Gamma(V\rightarrow e^{+}e^{-})$,
then the unknown free parameters $({f^{(i)}_{\omega NN}}/{f^{s}_{\omega}}),
({f^{(i)}_{\phi NN}}/{f^{s}_{\phi}})$ of a strange nucleon FF's model

\begin{equation}
  F^{s}_{i}(t)=\bar{f}\left[t;({f^{(i)}_{\omega NN}}/{f^{s}_{\omega}}),
({f^{(i)}_{\phi NN}}/{f^{s}_{\phi}})\right] (i=1,2)
\label{Ms}
\end{equation}
of the same inner analytic structure as the isoscalar parts of the nucleon
EM FF's, but of course with different norms and possibly
with different asymptotics (therefore denoted by $\bar{f}$), are
numerically evaluated by the relations \cite{jaf}

\begin{eqnarray}
\nonumber
({f^{(i)}_{\omega NN}}/{f^{s}_{\omega}})&=&-\sqrt{6}\frac{\sin{\varepsilon}}
{\sin(\varepsilon+\theta_{0})}({f^{(i)}_{\omega NN}}/{f^{e}_{\omega}})\\
({f^{(i)}_{\phi NN}}/{f^{s}_{\phi}})
&=&-\sqrt{6}\frac{\cos{\varepsilon}}
{\cos(\varepsilon+\theta_{0})}({f^{(i)}_{\phi NN}}/{f^{e}_{\phi}})
\label{CC}\\
\nonumber
(i&=&1,2)
\end{eqnarray}
where $f^{s}_{\omega},f^{s}_{\phi}$ are strange-current $\leftrightarrow
V=\omega,\phi$ coupling constants and $\varepsilon=3.7^{0}$ is a deviation
from the ideally mixing angle $\theta_{0}=35.3^{0}$.

   The isoscalar Dirac and Pauli EM FF's of the nucleon
$F^{I=0}_{i}(t)$ (i=1,2), as well as the strange quark vector current Dirac
and Pauli FF's $F^{s}_{i}(t)$ (i=1,2), are assumed to be real analytic (i.e.
the reality condition $F^{\ast}_{i}(t)=F_{i}(t^{\ast})$ is satisfied) in the
whole complex t-plane except for cuts placed on the positive part of the real
axis between $t^{I=0}_{0}=9m^{2}_{\pi}$ ($m_{\pi}$ is the pion mass) and
$+\infty$.

   They are normalized at t=0

\begin{equation}
F^{I=0}_{1}(0)=\frac{1}{2}, \quad F^{I=0}_{2}(0)=\frac{1}{2}(\mu_{p}+\mu_{n}),
\label{N0}
\end{equation}
where $\mu_{p}$ and $\mu_{n}$ are the anomalous magnetic moments of the
proton and neutron, respectively. Similarly, as the overall strangeness
charge of the nucleon is zero, then

\begin{equation}
 F^{s}_{1}(0)=0, \quad F^{s}_{2}(0)=\mu_{s},
\label{NS}
\end{equation}
where $\mu_{s}$ is the strangeness nucleon magnetic moment, expected to be
determined numerically from a predicted behaviour of $F^{s}_{2}(t)$.

   At a very large space-like $t$ the $F^{I=0}_{1}(t)$ and $F^{I=0}_{2}(t)$
obey the asymptotic behaviour \cite{lep}

\begin{equation}
 F^{I=0}_{1}(t)_{|t|\rightarrow\infty}\sim t^{-2};\quad
 F^{I=0}_{2}(t)_{|t|\rightarrow\infty}\sim t^{-3}.
\label{AS0}
\end{equation}
For $F^{s}_{1}(t)$ and $F^{s}_{2}(t)$ we assume the same asymptotic
conditions to be fulfilled.

   All these properties, together with the experimental fact of a creation of
vector-meson resonances of the photon quantum numbers in electron-positron
annihilation processes into  hadrons, are contained consistently in the
following unitary and analytic models \cite{{dnc},{dzp}}

\begin{eqnarray}
&& F^{I=0}_{1}[v(t)]
 =(\frac{1-v^{2}}{1-v^{2}_{N}})^{4}\{\frac{1}{2}
 L(v_{\omega''})L(v_{\omega'})+\nonumber\\
&& [L(v_{\omega''})L(v_{\omega})
 \frac{(C_{\omega''}-C_{\omega})}
 {(C_{\omega''}-C_{\omega'})}-
 L(v_{\omega'})L(v_{\omega})
 \frac{(C_{\omega'}-C_{\omega})}
 {(C_{\omega''}-C_{\omega'})}\nonumber\\
&& -L(v_{\omega''})L(v_{\omega'})]
 (f^{(1)}_{\omega NN}/f^{e}_{\omega})+\nonumber\\
&& [L(v_{\omega''})L(v_{\phi})
 \frac{(C_{\omega''}-C_{\phi})}
 {(C_{\omega''}-C_{\omega'})}-
 L(v_{\omega'})L(v_{\phi})
 \frac{(C_{\omega'}-C_{\phi})}
 {(C_{\omega''}-C_{\omega'})}\nonumber\\
&& -L(v_{\omega''})L(v_{\omega'})]
 (f^{(1)}_{\phi NN}/f^{e}_{\phi})\},
\label{F01}
\end{eqnarray}

\begin{eqnarray}
\nonumber
&& F^{I=0}_{2}[v(t)]=(\frac{1-v^2}{1-v^2_N})^6\{L(v_{\omega''})
 L(v_{\omega'})L(v_{\omega})\\
\nonumber
&& [1-{\frac{C_{\omega}}{(C_{\omega''}-C_{\omega'})}}
 (\frac{(C_{\omega''}-C_{\omega})}{C_{\omega'}}-
 \frac{(C_{\omega'}-C_{\omega})}{C_{\omega''}})]\\
\nonumber
&& (f^{(2)}_{\omega NN}/f^{e}_{\omega})+
 L(v_{\omega''})L(v_{\omega'})L(v_{\phi})\\
\nonumber
&& [1-{\frac{C_{\phi}}{(C_{\omega''}-C_{\omega'})}}
 (\frac{(C_{\omega''}-C_{\phi})}{C_{\omega'}}-
 \frac{(C_{\omega'}-C_{\phi})}{C_{\omega''}})]\\
&& (f^{(2)}_{\phi NN}/f^{e}_{\phi})\}
\label{F02}
\end{eqnarray}
and

\begin{eqnarray}
&& F^{s}_{1}[v(t)]=(\frac{1-v^{2}}{1-v^{2}_{N}})^{4}\times\nonumber\\
&& \{[L(v_{\omega''})L(v_{\omega})
 \frac{(C_{\omega''}-C_{\omega})}
 {(C_{\omega''}-C_{\omega'})}-
 L(v_{\omega'})L(v_{\omega})
 \frac{(C_{\omega'}-C_{\omega})}
 {(C_{\omega''}-C_{\omega'})}\nonumber\\
&& -L(v_{\omega''})L(v_{\omega'})]
 (f^{(1)}_{\omega NN}/f^{s}_{\omega})+\nonumber\\
&& [L(v_{\omega''})L(v_{\phi})
 \frac{(C_{\omega''}-C_{\phi})}
 {(C_{\omega''}-C_{\omega'})}-
 L(v_{\omega'})L(v_{\phi})
 \frac{(C_{\omega'}-C_{\phi})}
 {(C_{\omega''}-C_{\omega'})}\nonumber\\
&& -L(v_{\omega''})L(v_{\omega'})]
 (f^{(1)}_{\phi NN}/f^{s}_{\phi})\}.
\label{FS1}
\end{eqnarray}
$F^{s}_{2}[v(t)]$ has the same analytic form as (\ref{F02}) with the
replacement $f^e_{\omega},f^e_{\phi} \rightarrow f^s_{\omega},f^s_{\phi}$.
All models are defined on a four-sheeted Riemann surface with complex
conjugate pairs of resonance poles placed only on the unphysical sheets,
where

\begin{eqnarray*}
\nonumber
&& L(v_r)=\frac{(v_N-v_r)(v_N-v^{\ast}_r)(v_N-1/v_r)(v_N-1/v^{\ast}_r)}
 {(v-v_r)(v-v^{\ast}_r)(v-1/v_r)(v-1/v^{\ast}_r)},\\
&& (r=\omega,\phi,\omega',\omega'')\\
\nonumber
&& C_r=\frac{(v_N-v_r)(v_N-v^{\ast}_r)(v_N-1/v_r)(v_N-1/v^{\ast}_r)}
 {-(v_r-1/v_r)(v^{\ast}_r-1/v^{\ast}_r)},\\
&& (r=\omega,\phi,\omega',\omega'')
\end{eqnarray*}
\end{multicols}

\begin{equation}
 v(t)=i\frac
 {\sqrt{[\frac{t_{N\bar N}-t^{I=0}_0}{t^{I=0}_0}]^{1/2}+
 [\frac{t-t^{I=0}_0}{t^{I=0}_0}]^{1/2}}-
 \sqrt{[\frac{t_{N\bar N}-t^{I=0}_0}{t^{I=0}_0}]^{1/2}-
 [\frac{t-t^{I=0}_0}{t^{I=0}_0}]^{1/2}}}
 {\sqrt{[\frac{t_{N\bar N}-t^{I=0}_0}{t^{I=0}_0}]^{1/2}+
 [\frac{t-t^{I=0}_0}{t^{I=0}_0}]^{1/2}}+
 \sqrt{[\frac{t_{N\bar N}-t^{I=0}_0}{t^{I=0}_0}]^{1/2}-
 [\frac{t-t^{I=0}_0}{t^{I=0}_0}]^{1/2}}}
\label{ITR}
\end{equation}

\begin{equation}
 v_N=v(t)_{|t=0}; v_r=v(t)_{|t=(m_r-i\Gamma_r/2)^2};
 (r=\omega,\phi,\omega',\omega''),
\label{DEF}
\end{equation}

\begin{multicols}{2}
and $t_{N\bar N}=4m^2_N$ is a square-root branch point corresponding to $N\bar N$
threshold.

   Here we would like to emphasize the following two facts concerning
the unitary and analytic models of $F^{I=0}_1[v(t)]$, $F^{I=0}_2[v(t)]$,
$F^s_1[v(t)]$, $F^s_2[v(t)]$.

   Since $\mu_s$ is unknown in advance, one could not utilize the second
normalization of (\ref{NS}) in a construction of the model for $F^s_2[v(t)]$.
In order to keep the same inner analytic structure of $F^{I=0}_2[v(t)]$,
we did not apply neither the second normalization condition of (\ref{N0}) in a
construction of the model (\ref{F02}).

   In despite of the saturation of $F^{I=0}_i[v(t)]$ and $F^s_i[v(t)]$,
$(i=1,2)$ by four isoscalar vector-mesons $\omega,\phi, \omega', \omega''$
those FF's depend finally only on $\omega$ and $\phi$ coupling
constant ratios. This reduction of the number of free parameters is
a consequence of an application of normalizations (\ref{N0}),(\ref{NS})
and the asymptotic conditions (\ref{AS0}) to the constructed models, which
enabled us to express the coupling constant ratios of $\omega'$ and $\omega''$
through $\omega$ and $\phi$ coupling constant ratios.

   The latter in $F^{I=0}_1[v(t)]$ and $F^{I=0}_2[v(t)]$ can be evaluated
numerically by a comparison of expressions (\ref{F01}) and (\ref{F02}) with
experimental data on proton and neutron electric and magnetic FF's,
$G^p_E(t), G^p_M(t), G^n_E(t), G^n_M(t),$ in the space-like and also (this is
crucial) in the time-like regions simultaneously. However,
in a realization of such a program we are in need also of the unitary and
analytic models of the isovector parts of the Dirac and Pauli nucleon FF's:

\begin{eqnarray}
\nonumber
&& F^{I=1}_{1}[w(t)]
 =(\frac{1-w^{2}}{1-w^{2}_{N}})^{4}\{\frac{1}{2}
 H(w_{\rho'''})L(w_{\rho''})+\\
\nonumber
&& [H(w_{\rho'''})L(w_{\rho})
 \frac{(D_{\rho'''}-D_{\rho})}
 {(D_{\rho'''}-D_{\rho''})}-
 L(w_{\rho''})L(w_{\rho})\\
\nonumber
&& \frac{(D_{\rho''}-D_{\rho})}
 {(D_{\rho'''}-D_{\rho''})}-
 H(w_{\rho'''})L(w_{\rho''})]
 (f^{(1)}_{\rho NN}/f^{e}_{\rho})+\\
\nonumber
&& [H(w_{\rho'''})L(w_{\rho'})
 \frac{(D_{\rho'''}-D_{\rho'})}
 {(D_{\rho'''}-D_{\rho''})}-
 L(w_{\rho''})L(w_{\rho'})\\
&& \frac{(D_{\rho''}-D_{\rho'})}
 {(D_{\rho'''}-D_{\rho''})}-
 H(w_{\rho'''})L(w_{\rho''})]
 (f^{(1)}_{\rho' NN}/f^{e}_{\rho'})\},
\label{F11}
\end{eqnarray}

\begin{eqnarray}
&& F^{I=1}_{2}[w(t)]
 =(\frac{1-w^{2}}{1-w^{2}_{N}})^{6}\times\nonumber\\
&& \{\frac{1}{2}(\mu_p+\mu_n)
 H(w_{\rho'''})L(w_{\rho''})L(w_{\rho'})+\nonumber\\
&& [H(w_{\rho'''})L(w_{\rho''})L(w_{\rho})
 \frac{(D_{\rho'''}-D_{\rho})
 (D_{\rho''}-D_{\rho})}
 {(D_{\rho'''}-D_{\rho'})
 (D_{\rho''}-D_{\rho'})}\nonumber\\
&& +L(w_{\rho''})L(w_{\rho'})L(w_{\rho})
 \frac{(D_{\rho''}-D_{\rho})
 (D_{\rho'}-D_{\rho})}
 {(D_{\rho'''}-D_{\rho''})
 (D_{\rho'''}-D_{\rho'})}\nonumber\\
&& -H(w_{\rho'''})L(w_{\rho'})L(w_{\rho})
 \frac{(D_{\rho'''}-D_{\rho})
 (D_{\rho'}-D_{\rho})}
 {(D_{\rho'''}-D_{\rho''})
 (D_{\rho''}-D_{\rho'})}\nonumber\\
&& -H(w_{\rho'''})L(w_{\rho''})L(w_{\rho'})]
 (f^{(2)}_{\rho NN}/f^{e}_{\rho})\}\\
\label{F21}
\nonumber
\end{eqnarray}
with

\begin{eqnarray}
\nonumber
&& H(w_v)=\frac{(w_N-w_v)
 (w_N-w^{\ast}_v)(w_N+w_v)
 (w_N+w^{\ast}_v)}{(w-w_v)
 (w-w^{\ast}_v)(w+w_v)
 (w+w^{\ast}_v)},\\
\nonumber
\end{eqnarray}
as there are relations between $G^{p,n}_E(t), G^{p,n}_M(t)$ and
$F^{I=0}_i(t), F^{I=1}_i(t)$ $(i=1,2)$ as follows

\begin{eqnarray}
\nonumber
&& G^{p,n}_E=[F^{I=0}_1 \pm F^{I=1}_1]+\frac{t}{4m^2_{p,n}}
 [F^{I=0}_2 \pm F^{I=1}_2]\\
&& G^{p,n}_M=[F^{I=0}_1 \pm F^{I=1}_1]+
 [F^{I=0}_2 \pm F^{I=1}_2].
\label{GFF}
\end{eqnarray}
The optimal description of all existing data on $G^{p,n}_{E,M}(t)$ is
achieved with the following numerical values of the corresponding
coupling constant ratios

\begin{eqnarray}
\nonumber
&& (f^{(1)}_{\omega NN}/f^e_{\omega})=0.73045; \quad
 (f^{(1)}_{\phi NN}/f^e_{\phi})=0.12139;\\
\nonumber
&& (f^{(2)}_{\omega NN}/f^e_{\omega})=0.20178; \quad
 (f^{(2)}_{\phi NN}/f^e_{\phi})=-.25619; \\
\nonumber
&& (f^{(1)}_{\rho NN}/f^e_{\rho})=0.51613; \quad
 (f^{(1)}_{\rho' NN}/f^e_{\rho'})=0.71113;\\
&& (f^{(2)}_{\rho NN}/f^e_{\rho})=3.3377,
\label{ECC}
\end{eqnarray}
provided all masses and widths of vector-mesons
under considerations (except for the $\rho'''(2150)$, for which they are
taken from \cite{bnc}) are fixed at the table values \cite{rpp}.
Then from (\ref{ECC}), by means of the relations (\ref{CC}), one finds

\begin{eqnarray}
\nonumber
&& (f^{(1)}_{\omega NN}/f^s_{\omega})=-.18347; \quad
 (f^{(1)}_{\phi NN}/f^s_{\phi})=-.38181;\\
&& (f^{(2)}_{\omega NN}/f^s_{\omega})=-.05068; \quad
 (f^{(2)}_{\phi NN}/f^s_{\phi})=0.80580.
\label{SCC}
\end{eqnarray}
Substituting the latter into $F^s_1[v(t)]$ and $F^s_2[v(t)]$ and defining
strange nucleon electric and magnetic FF's

\begin{eqnarray}
\nonumber
&& G^s_E(t)=F^s_1(t)+ \frac{t}{4m^2_N}F^s_2(t)\\
&& G^s_M(t)=F^s_1(t)+F^s_2(t)
\end{eqnarray}
in analogy to nucleon Sachs FF's, one predicts the behaviour of $G^s_E(t)$ and
$G^s_M(t)$ as they are presented in Figs. 1 and 2, respectively.

\begin{figure}
\centering
\epsfxsize=0.45\textwidth\epsffile{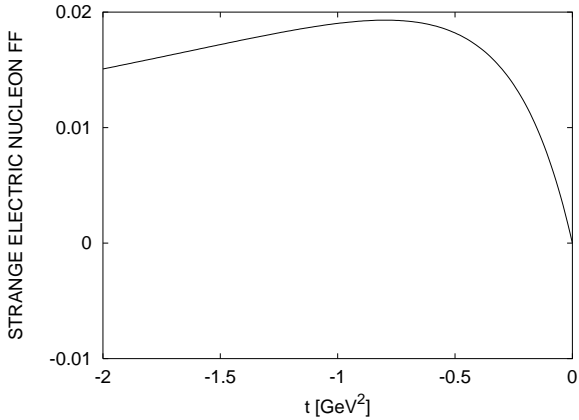}
\caption{The predicted behaviour of $G^s_E(t)$.}
\end{figure}

\begin{figure}
\centering
\epsfxsize=0.45\textwidth\epsffile{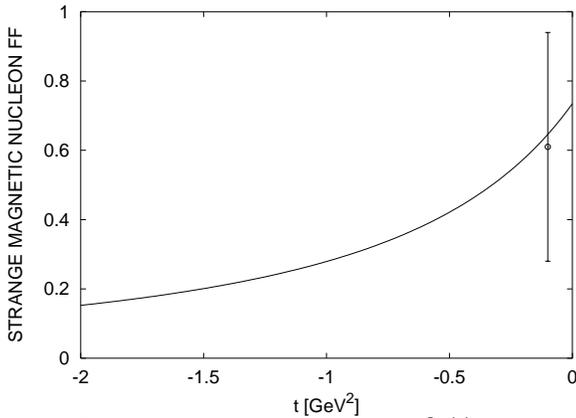}
\caption{The predicted behaviour of $G^s_M(t)$ and its comparison
with the improved SAMPLE experimental result (3).}
\end{figure}

One can see from Fig. 2 that the improved SAMPLE collaboration result
\cite{spa} for the central value is preferred to the older one \cite{mue} and
that the strange magnetic moment of the nucleon takes the value
$\mu_s=G^s_M(0)=+0.73 [\mu_N]$.

   On the other hand, by using the same behaviour of $G^s_E(t)$ and $G^s_M(t)$
one can predict at $t=-0.48 GeV^2$ a value of the combination
\begin{equation}
G^s_E(-0.48)+0.39G^s_M(-0.48)=+0.185\\
\label{CEM}
\end{equation}
again indicating the positive central value like in (\ref{GEM}).

   This behaviour gives also a prediction at $t=-0.23 GeV^2$
for the following combination
\begin{equation}
G^s_E(-0.23)+0.22G^s_M(-0.23)=+0.135;
\end{equation}
just measured at the MAMI A4 \cite{har} running experiment.

   In conclusion, we would like to emphasize that all these predictions
supporting the positive central values on the strange nucleon FF's from
recent SAMPLE and HAPPEX experiments, are based on two key ingredients.
Form factor models incorporate assumed analytic properties and as a result
the simultaneous description of all available data on the nucleon EM FF's,
including also the time-like region, is achieved.

\medskip

   The work was in part supported by Slovak Grant Agency for Sciences, Gr. No
2/5085/2000 (S.D.) and Gr. No 1/7068/2000 (A.Z.D.).

\end{multicols}

\end{document}